\documentclass[12pt] {article}
\pdfoutput=1
\usepackage[hyphens,spaces,obeyspaces]{url}
\usepackage{setspace}

\usepackage{epsfig}
\usepackage{amssymb}
\usepackage{amsmath}
\usepackage{multirow}
\usepackage{setspace}
\usepackage{datetime}
\usepackage{amsmath}
\usepackage{algorithm}
\usepackage{algpseudocode}
\usepackage{graphicx}
\usepackage{lipsum}
\usepackage{acronym}
\usepackage{array}
\usepackage{graphicx}
\usepackage{cite}
\usepackage{graphicx}
\usepackage{epstopdf}
\usepackage{caption}
\usepackage{balance}
\usepackage{fancybox}
\usepackage{colortbl}
\usepackage{array}
\usepackage{mystyle}
\usepackage{multirow}
\usepackage[obeyspaces,hyphens,spaces]{url}
\usepackage{threeparttable}
\usepackage[english]{babel}% tables
\usepackage{longtable}
\usepackage{listings, lstautogobble}
\usepackage{booktabs}
\usepackage{threeparttable}
\usepackage[table,xcdraw]{xcolor}
\usepackage{wrapfig}
\usepackage{subfigure}
\usepackage{enumitem}
\usepackage{paralist}

\usepackage{amsthm}
\AtBeginEnvironment{thm}{\footnotesize}
\AtBeginEnvironment{defn}{\footnotesize}
\providecommand{\keywords}[1]{\textbf{\textit{Keywords---}} #1}

\usepackage{stackengine}
\newsavebox\mybox

\newtheorem{definition}{Definition}[section]

\newtheorem{theorem}{Theorem}[section]
\usepackage{hyperref}
\usepackage[framemethod=tikz]{mdframed}
\usepackage{enumitem}
\theoremstyle{definition}
\newtheorem{thm}{\small{Theorem}}%[section] % reset theorem numbering for each section
\AtBeginEnvironment{thm}{\footnotesize}
\AtBeginEnvironment{defn}{\footnotesize}

{\begin{list}{}%
         {\setlength{\leftmargin}{#1}}%
         \item[]%
}
{\end{list}}

\newdateformat{monthyeardate}{%
  \monthname[\THEMONTH], \THEYEAR}
\begin{document}
\sloppy

\title{\Large{\textbf{A Formally Verified HOL Algebra for Dynamic Reliability Block Diagrams} }}
\author{
Yassmeen~Elderhalli, Osman~Hasan, and Sofi\`ene~Tahar\vspace*{2em}\\
Department of Electrical and Computer Engineering,\\
Concordia University, Montr\'eal, QC, Canada 
\vspace*{1em}\\
\{y\_elderh,o\_hasan,tahar\}@ece.concordia.ca 
 \vspace*{3em}\\
\textbf{TECHNICAL REPORT}\\
\date{August 2019}
}
\maketitle

\newpage
\begin{abstract}

Dynamic reliability block diagrams (DRBDs) are introduced to overcome the modeling limitations of traditional reliability block diagrams, such as the inability to capture redundant components. However, so far there is no algebraic framework that allows conducting the analysis of a given DRBD based on its structure function and enables verifying its soundness using higher-order logic (HOL) theorem proving. In this work, we propose a new algebra to formally express the structure function and the reliability of a DRBD with spare constructs based on basic system blocks and newly introduced DRBD operators. We present several simplification properties that allow reducing the structure of a given DRBD. We provide the HOL formalization of the proposed algebra, and formally verify its corresponding properties using the HOL4 theorem prover. This includes formally verifying generic reliability expressions of the spare construct, series, parallel and deeper structures in an extensible manner that allows verifying the reliability of complex systems. Finally, we demonstrate the applicability of this algebra by formally analyzing the terminal reliability analysis of a shuffle-exchange network in HOL4.  
\end{abstract}

\section{Introduction}
\label{introduction}
%\vspace{-5pt}
Reliability measures how reliable a system is by finding the probability that it will continue to provide its desirable service in a given period of time. Fault trees (FTs) \cite{DFT-survey} and reliability block diagrams (RBDs) \cite{hasan2015reliability} are the most commonly used reliability modeling techniques. FTs graphically model the sources of failure of a system or subsystem using FT gates. An RBD, on the other hand, is a graphical representation of the reliability of a system. The components of a system are modeled as blocks and are connected using connectors (lines) to create a path or multiple paths from the RBD input to its output. These paths represent the required working blocks (system components) for the system to have a successful operation. The modeled system fails when components fail in such a manner that leads to the disconnection of all the paths between the input and the output. RBDs can be connected in a \textit{series}, \textit{parallel}, \textit{series-parallel} or \textit{parallel-series} fashion to create the appropriate modeling structure depending on the behavior and the components redundancy of the modeled system, which provides flexible and extensible modeling configurations to represent complex systems. However, both of the traditional RBDs and FTs are unable to model the dynamic behavior of system components, where the change of a state of one component can affect the state of other system components.\\    
\indent Dynamic fault trees (DFTs) \cite{DFT-survey} are proposed as an extension to traditional FTs by introducing DFT gates, such as spare gates. However, the only behavior that is captured by DFTs is the dynamic failure effect of one system component in the failure or activation of other components. To overcome the modeling limitations of DFTs, RBDs are extended to \textit{dynamic reliability block diagrams} (DRBDs) to model the dynamic dependency among system components in several scenarios by introducing new constructs \cite{Distefano-Thesis}. These new constructs are basically DRBD blocks that  enable modeling dynamic relationships among system components, such as a load sharing construct that captures the effect of sharing a load on the reliability of system components and spare construct that models the reliability of spare parts in a DRBD. \\
\indent Formal methods have been used in the analysis of RBDs; both the dynamic and static (traditional) ones. For instance, in \cite{xu2007formal}, the formal semantics of DRBD constructs in Object-Z formalism \cite{smith2012object} are proposed. However, analyzing and verifying the behavior of DRBDs based on this formalism is not feasible since there is a lack in the support of the used tools. Therefore, in \cite{smith2012object}, the DRBDs are then converted into a Colored Petri Net (CPN) to be analyzed using Petri nets tools. An algorithm to automatically convert a DRBD into a CPN is also proposed in \cite{robidoux2010automated}. Since CPN is used, only some state-based properties of the modeled system can be analyzed. In \cite{ahmed2016formalization}, the HOL4 theorem prover \cite{HOL4} is used to formalize several configurations of static RBDs. However, this formalization can only analyze the combinatorial behavior of systems and cannot provide support to analyze DRBDs. In addition, this formalization cannot be tailored to provide support for DRBDs, and thus, it is required to have a brand new higher-order logic (HOL) formalization to support this kind of analysis. \\

\indent In system engineering, it is important to be able to analyze DRBDs qualitatively in order to identify the sources of system vulnerability, and quantitatively in order to evaluate the system reliability. However, to the best of our knowledge, so far there exists  no algebraic approach that mathematically models a given DRBD and enables expressing its function based on basic components just like the DFT algebra \cite{Merle-thesis}. Using such algebra in the reliability analysis will result in simpler and fewer proof steps than the DFT-based algebraic analysis \cite{Merle-thesis}, since the probabilistic principle of inclusion and exclusion will not be invoked. In this work, we propose a new algebraic approach for DRBD analysis that allows having a DRBD expression to be used for both qualitative and quantitative analyses.  
We introduce new operators to mathematically model the dynamic behavior in DRBD structures and constructs. In particular, we use these operators to model a DRBD spare construct besides traditional series, parallel, series-parallel and parallel-series structures. Moreover, we provide simplification theorems that allow reducing the structure of a given DRBD. This DRBD structure can be then analyzed to obtain a generic expression of the system reliability. The reliability expressions obtained using this approach are generic and independent of the distribution and density functions that represent the system components. Although basic operators, such as OR and AND, were introduced in \cite{Distefano-Thesis}, they are only useful to model parallel and series constructs of dependent components. Moreover, there is no general mathematical expression that would allow reasoning about the behavior of DRBDs. In addition, the DRBDs constructs of \cite{Distefano-Thesis} are quite complex, which complicates  modeling large systems. Therefore, we use the constructs proposed in \cite{xu2007formal} as they are much simpler, which facilitates defining the new algebra to model various new DRBD constructs. In this work, we use the DRBD constructs of \cite{xu2007formal}. Leveraging upon the expressive nature of HOL, we formally verify the soundness of the proposed DRBD algebra using HOL theorem proving. Although the formalization development can be conducted using many theorem provers, we choose the HOL4 theorem prover, as our existing formalization of DFT algebra can be useful since our proposed DRBD algebra is compatible with the DFT's. The work contributions can be summarized as: 

\begin{compactitem}
\item  A new DRBD algebra that includes DRBD operators and simplification theorems that allow expressing the structure of a given DRBD.
\item  A HOL formalization of the introduced DRBD algebra, i.e., modeling the DRBD operators and verifying their simplification theorems using HOL4 to ensure the soundness of the proposed approach.  
\item  A mathematical expression and HOL formalization of the spare construct and its reliability.
\item Mathematical models and reliability expressions of the traditional series, parallel and deeper structures for an arbitrary number of inputs using the new DRBD operators with their HOL formalization. 
\item Formal reliability analysis of two real-world systems
\end{compactitem}

Our ultimate goal is to develop a formally verified algebra that follows the traditional reliability expressions of the series and parallel structures in an easily extensible manner and at the same time can capture the dynamic behavior of real-world systems. 
Our formalization differs from and overcomes the formalization of traditional RBDs presented in \cite{ahmed2016formalization} in the sense that it can formally express the structure function of a DRBD using the introduced DRBD operators. In addition, it can formally model and analyze DRBD spare constructs. Furthermore, we model the traditional RBD structures, i.e., series, parallel and deeper structures in a way similar to the mathematical models available in the literature, which makes it easily understood and followed by reliability engineers that are not familiar with HOL theorem proving. Finally, we illustrate the usefulness of the proposed developments in conducting the formal analysis of two real-world systems; the terminal reliability of a shuffle-exchange network and the reliability of a drive-by-wire system.

\section{DRBD Algebra}
\label{algebra}
In this section, we present the proposed algebra for DRBD analysis. This algebra allows modeling the structure function of DRBDs with spare constructs. Moreover, we present some simplification properties that enable reducing the structure function when possible. Throughout this work, we assume that system components or blocks are represented by random variables that in turn represent their time-to-failures. In addition, we assume that system components are non-repairable, i.e., we are interested in expressing the reliability of the system considering that the failed components will not be repaired. It is worth mentioning that our proposed algebra follows the general lines for the DFT algebra \cite{Merle-thesis}, which allows DFTs conversion into DRBDs for conducting their analysis as well. 

The reliability of a single component, which time-to-failure function is represented by random variable $X$, is mathematically defined as \cite{hasan2015reliability}:
%\vspace{-3pt}
\begin{equation}
\label{eq:rel}
R_{X}(t)=Pr\{s\ |\ X(s)\ >\ t\} = 1-Pr \{s\ |\ X(s)\ \leq\ t\} = 1-F_{X}(t)
\end{equation}
%\vspace{-5pt}
\noindent where $F_{X}(t)$ is the cumulative distribution function (CDF) of $X$.

We call $\{s\ |\ X(s)\ >\ t\}$ as a DRBD event as it represents the set that we are interested in finding the probability of until time $t$:
%\vspace{-3pt}
\begin{equation}
\label{eq:event}
event\ (X,\ t)\ =\  \{s\ |\ X(s)\ >\ t\}
\end{equation}
%\vspace{-30pt}
\subsection{Identity Elements, Operators and Simplification Properties}
\begin{figure}[!t]
%\vspace{-10pt}
\subfigure[Series DRBD]{
   \makebox[0.5\textwidth]{

{\includegraphics[scale=0.7]{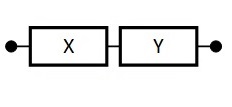}}}} 
\hfill
\subfigure[Parallel DRBD]{
  \makebox[0.5\textwidth]{
  {\includegraphics[scale=0.7]{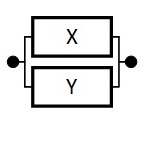}
}}}
\caption{Two-Block Series and Parallel DRBDs}\label{fig:two-block-DRBD}

\end{figure}

Similar to the identity elements of ordinary Boolean algebra and DFT algebra \cite{Merle-thesis}, we introduce two identity elements, i.e., ALWAYS and NEVER, that represent two states of any system block.  The \textit{ALWAYS} element represents a system component that always fails, i.e., it fails from time $0$. While the \textit{NEVER} element represents a component that never fails, i.e., the time of its failure is $+\infty$. These identity elements play an important role in the reduction process of the structure functions of DRBDs, as will be introduced in the following sections. 
%\vspace{-5pt}
\begin{equation}
ALWAYS = 0
\end{equation}
%\vspace{-24pt}
\begin{equation}
NEVER = +\infty
\end{equation}
We introduce operators to model the relationship between the various blocks in a DRBD. These operators can be divided into two categories: 1) The AND and OR operators that are not concerned with the dependencies among system components. 2) Temporal operators, i.e., \textit{After}, \textit{Simultaneous} and \textit{Inclusive After}, that can capture the dependencies between system components. It is worth mentioning that DRBDs are concerned with modeling the several paths of success of a given system. Therefore, if we are concerned in knowing the success behavior of a DRBD until time $t$, it means that we are interested in knowing how the system would not fail until time $t$. As a result, we can use the time-to-failure random variables in modeling the time-to-failure of a given DRBD, i.e., its structure function. It is assumed that for any two system components that possess continuous failure distribution functions, the possibility that these components fail at the same time can be neglected. \\
\indent In \cite{Distefano-Thesis}, AND and OR operators were introduced to model the parallel and series constructs between dependent components only without providing any mathematical model to these operators. We propose to use the AND ($\cdot$) and OR ($+$) operators to model series and parallel blocks in a DRBD, respectively without any restriction. We provide a mathematical model for each operator based on the time of failure of its inputs, as listed in Table~\ref{table:or_and_reliability}, to be used in the proposed algebra. The AND operator models the series connection between two or more system blocks, as shown in Figure~\ref{fig:two-block-DRBD}(a). For example, the DRBD in Figure~\ref{fig:two-block-DRBD}(a) will continue to work only if component $X$ and component $Y$ are working. Once one of these blocks stops working, then there will be no connection between the input and the output of the DRBD and thus the system will no longer work. We model the AND operator as the minimum time of its input arguments. Similarly, the OR operator models the connection between parallel components in a DRBD. For example, the DRBD in Figure~\ref{fig:two-block-DRBD}(b) will continue to work if $X$ is working or $Y$ is working. All the components in a parallel structure should fail for this DRBD to fail. Therefore, we model the OR operator as the maximum time of failure of its input arguments, which represents the time of failure of basic system blocks or sub-DRBDs. This approach facilitates using these operators to model even the complex structures.    

\begin{table}[b]
\centering
%\vspace{-20pt}
\caption{Mathematical and Reliability Expressions of AND and OR Operators}
%\vspace{-5pt}
\label{table:or_and_reliability}
\renewcommand{\arraystretch}{1.2}
\begin{tabular}{|c|l|l|}
\hline
{Operator} & \multicolumn{1}{c|}{{Math. Expression}} & \multicolumn{1}{c|}{{Reliability}} \\ \hline \hline
{AND}      &   $ X \cdot Y= min\ (X,Y)$                                    & $ R_{(X\cdot Y)}(t)\ = R_{X}(t)\ \times\ R_{Y}(t)$                                \\ \hline
{OR}       &  $ X + Y = max\ (X,Y)$                                     &  $  R_{(X+ Y)}(t)\ = 1- ((1-R_{X}(t))\times(1-R_{Y}(t)))$                                \\ \hline
\end{tabular}
\end{table}
If $X$ and $Y$ are independent, then the reliability of the systems, shown in Figure \ref{fig:two-block-DRBD}, can be expressed as in Table \ref{table:or_and_reliability}. To reach these expressions, it is required first to express the DRBD events as the intersection and union for the AND and OR operators, respectively, as:
%\vspace{-5pt}
\begin{equation}
\label{and-intersect}
event\ ((X\cdot Y),\ t)\ =\  event\ (X,\ t)\ \cap\ event\ (Y,\ t)
\end{equation} 
%\vspace{-20pt}
\begin{equation}
\label{or-union}
event\ ((X+ Y),\ t)\ =\  event\ (X,\ t)\ \cup\ event\ (Y,\ t)
\end{equation} 
In order to model the dynamic behavior of systems in DRBDs, we introduce new temporal operators: \textit{after} ($\rhd$), \textit{simultaneous} ($\Delta$), and \textit{inclusive after}($\unrhd$), as listed in Table~\ref{table:temporal_operators}. The \textit{after} operator represents a situation where it is required to model a component that continues to work after the failure of another. The time of failure of the after operator equals the time of failure of the last component, which is required to fail. However, if the required sequence does not occur, then the output can never fail, i.e., the time of failure equals $+\infty$.
\begin{table}[t]
%\vspace{-15pt}
\caption{Mathematical Expressions of Temporal Operators}
%\vspace{-5pt}
\label{table:temporal_operators}
\centering
\begin{tabular}{|c||c||c|}
\hline
 {After($\rhd$)}&{Simultaneous($\Delta$)}  & {Inclusive After($\unrhd$)}\\ \hline \hline
$ X  \rhd Y =   \begin{cases}  X, & X > Y\\[-1\jot]  +\infty,  &  X  \leq Y 
\end{cases}$ & $  X  \Delta Y =  \begin{cases}  X, & X =   Y\\[-1\jot]  +\infty,  & X  \neq Y 
\end{cases} $ & $  X  \unrhd Y =  \begin{cases}  X, & X \geq Y\\[-1\jot] + \infty,  & X  < Y 
\end{cases} $  \\ \hline
\end{tabular}
%\vspace{-5pt}
\end{table}
The behavior of the simultaneous operator is similar to the one introduced in the DFT algebra \cite{Merle-thesis}. The output of this operator fails if both its inputs fail at the same time, otherwise it can never fail.
Finally, the inclusive after operator encompasses the behavior of both the after and simultaneous operators, i.e, it models a situation where it is required that one component continues to work after another one or fail at the same time, otherwise it can never fail. In the case of dealing with basic components, the inclusive after will behave in a similar way as the after operator. Therefore, their probabilities can be expressed for independent random variables in the same way as: 
%\vspace{-5pt}
\begin{equation}
R_{(X \rhd Y)}(t) = 1-\int_{0}^{t} f_{X}(x) \times F_{Y}(x)\ dx
\end{equation}
\noindent where $F_{X}$ is the probability density function (PDF) of $X$ and $F_Y$ is the CDF of $Y$.

We introduce several simplification properties to reduce the structure function of a DRBD. These simplification properties range from simple ones, such as the associativity and idempotence of the operators, to more complex theorems. The idea of these properties is to reduce the algebraic expressions based on the time of failure. For example, $X \cdot ALWAYS = ALWAYS$ means that if a component in a series structure is not working, i.e., always fails, then the series structure is not working as well. Similarly, $X + NEVER = NEVER$ means that if a component in a parallel structure cannot fail, then the whole parallel structure cannot fail as well. 
$X+Y = Y+X$, $X\cdot Y = Y \cdot X$ and $X\Delta Y = Y \Delta X$ represent the commutativity property for the OR, AND and simultaneous operators, respectively. An example of a more complex theorem is $X \rhd (Y\cdot Z) = (X \rhd Y)\cdot( X \rhd Z)$. In Section \ref{Formalization_in_hol}, a full list of the developed theorems will be introduced. 

\subsection{DRBD Constructs and Structures}
The spare construct, shown in Figure \ref{fig:spare} \cite{xu2007formal}, is introduced in DRBDs to model situations where a spare part is activated and replaces the main part, after its failure, by introducing a spare controller to activate the spare  \cite{xu2007formal}. Depending on the failure behavior of the spare part, we can have three variants, i.e., hot, warm and cold ($H|W|C$) spares. The hot spare possesses the same failure behavior in both its active and dormant states. The cold spare cannot fail in its dormant state and is only activated after the failure of the main part. The failure behavior of the warm spare in the dormant state is attenuated by a dormancy factor from the active state. In order to distinguish between the dormant and active states of the spare, just like the DFT algebra \cite{Merle-thesis}, we use two different symbols to model the spare part of the DRBD spare construct, one for the dormant state and the other for the active one. For the spare construct of Figure~\ref{fig:spare}, the spare $X$ is represented by $X_{a}$ and $X_{d}$ for the active and dormant states, respectively. After the failure ($F$) of the main part $Y$, $X$ will be activated ($A$) by the spare controller. We model the structure function of the spare construct ($Q_{spare}$) using the DRBD operators based on the description of its behavior as: 
\begin{figure}[b]
    \centering
%\vspace{-5pt}
    \includegraphics[scale=0.7]{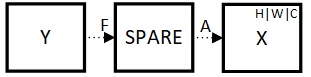}
%    \vspace{-10pt}
    \caption{Spare Construct}
    \label{fig:spare}
%\vspace{-10pt}
\end{figure}
\begin{equation}
\label{spare}
 Q_{spare}= (X_{a} \rhd Y)\cdot (Y \rhd X_{d}) 
\end{equation}   
Thus, we need two conditions to be satisfied in order for the spare to work. The first one is that the active state of the spare will continue to work after the failure of the main part $(X_{a}\rhd Y)$. The second condition is that the main part will continue to work after the failure of the spare in its dormant state $(Y \rhd X_{d})$. However, since the spare part can only fail in one of its states ($X_{a}, X_{d}$) but not both as it is non-repairable, only one of the terms in Equation~(\ref{spare}) affects the behavior and the other term can never fail, i.e., it fails at $+\infty$. 

Since the spare construct of the DRBD and the spare gate of the DFT exhibit complementary behavior, i.e., the DRBDs consider the success and the DFTs consider the failure, we can use the probability of failure of the spare DFT gate \cite{Merle-thesis} to find the reliability of the spare DRBD construct. It is assumed that the dormant spare and the main part are independent since the failure of one does not affect the failure of the other. However, the failure of the active spare is affected by the time of failure of the main part, since it will be activated after the failure of the main part. We express the reliability of the spare as: 
%\vspace{-3pt}
\begin{equation}
\label{spare_prob}
\begin{split}
R_{spare}(t) = 1 - \int_{0}^{t} \int_{y}^{t} f_{(X_{a}|Y=y)}(x)\ f_{Y}(y) dx dy - \int_{0}^{t} f_{Y}(y)F_{X_{d}}(y)dy
\end{split}
\end{equation}
\noindent where $f_{(X_{a}|Y=y)}$ is the conditional density function of $X_{a}$ given that $Y$ failed at time $y$.
Equations (\ref{spare}) and (\ref{spare_prob}) represent the general behavior of the spare, i.e., the warm spare. The hot and cold spares represent special cases of the warm spare and can be expressed as:
%\vspace{-3pt}
\begin{equation}
\label{hot_spare}
Q_{hot spare} = X + Y
\end{equation}
%\vspace{-20pt}
\begin{equation}
\label{cold_spare}
Q_{cold spare} = X_{a} \rhd Y
\end{equation}
In Equation (\ref{hot_spare}), the spare part $X$ has the same behavior in both states and thus there is no need to use any subscript to distinguish both states. The probability of Equation~(\ref{hot_spare}) can be expressed using the reliability of the OR operator, given in Table~\ref{table:or_and_reliability}. While the reliability of the cold spare construct can be expressed as:
%\vspace{-5pt}
\begin{equation}
R_{cold\ spare}(t) = 1 - \int_{0}^{t} \int_{y}^{t} f_{(X_{a}|Y=y)}(x)\ f_{Y}(y)\ dx\ dy
\end{equation}
%\vspace{-10pt}
\begin{table}[!b]
%%\vspace{-20pt}
\caption{Mathematical and Reliability Expressions of DRBD Structures}
%\vspace{-3pt}
\renewcommand{\arraystretch}{1.2}
\centering
\label{table_DRBD_structure}
\begin{tabular}{|c||c||c|}
\hline
{Structure}&{Math. Expression}  &{Reliability} \\ \hline\hline
{Series}& $ \bigcap_{i=1}^{n}(event\ (X_{i},\ t))$  &$ \prod_{i=1}^{n}R_{X_{i}}(t)$  \\ \hline
 {Parallel}& $ \bigcup_{i=1}^{n}(event\ (X_{i},\ t))$  &  $ 1- \prod_{1=1}^{n}(1-R_{X_{i}}(t))$\\ \hline
 {Series-Parallel} & $\bigcap_{i=1}^{m}\bigcup_{j=1}^{n}(event\ (X_{(i,j)},\ t))$ &  $ \prod_{i=1}^{m}(1- \prod_{j=1}^{n}(1-R_{X_{(i,j)}}(t)))$\\ \hline
{Parallel-Series} & $ \bigcup_{i=1}^{n}\bigcap_{j=1}^{m}(event\ (X_{(i,j)},\ t))$ & $  1-(\prod_{i=1}^{n}(1- \prod_{j=1}^{m}(R_{X_{(i,j)}}(t))))$\\ \hline
\end{tabular}
%\vspace{-10pt}
\end{table}

\begin{figure}[t]
\subfigure[Series]{
 \makebox[0.25\textwidth]{
\includegraphics[scale=0.6]{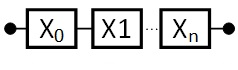}
}}
\subfigure[Parallel]{
 \makebox[0.2\textwidth]{
\includegraphics[scale=0.6]{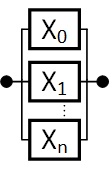}}}
\subfigure[Series-Parallel]{
 \makebox[0.25\textwidth]{
\includegraphics[scale=0.6]{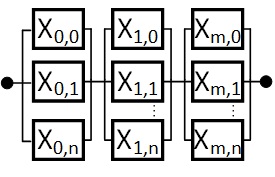}}}
\subfigure[Parallel-Series]{
 \makebox[0.2\textwidth]{
\includegraphics[scale=0.6]{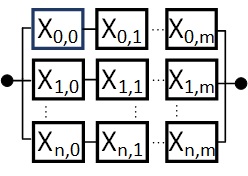}}}
%\vspace{-5pt}
\caption{DRBD Structures}
%\vspace{-12pt}
\label{fig:rbd_structures}
\end{figure}
%\vspace{-2pt}
Table \ref{table_DRBD_structure} lists the mathematical and reliability expressions of these structures \cite{hasan2015reliability}. The series structure represents a collection of blocks that are connected in series, as shown in Figure \ref{fig:rbd_structures}(a). The system continues to work until the failure of one of these blocks.  We define a series structure that represents the intersection of all events of the blocks in this structure as in Table \ref{table_DRBD_structure}, where $X_{i}$ represents the $i^{th}$ block in the series structure and $n$ is the number of blocks. Interestingly, any block in our proposed algebra can represent a basic system component or a complex structure, such as a spare construct. Moreover, since we are dealing with the events, we can use the ordinary reliability expressions for the series structure assuming the independence of the individual blocks. The parallel structure, shown in Figure \ref{fig:rbd_structures}(b), represents a system that continues to work until the failure of the last block in the structure. The behavior of the parallel structure can be expressed using the OR operator.  We represent the parallel structure as the union of the individual events of the blocks. The series-parallel structure, shown in Figure~\ref{fig:rbd_structures}(c), represents a series structure, where the blocks of the series structure are parallel structures. The structure function of this structure can be expressed using AND of ORs operators. Table~\ref{table_DRBD_structure} lists the model for this structure with its reliability expression, where $n$ is the number of blocks in the parallel structure and $m$ is the number of parallel structures that are connected in series. The parallel-series structure represents a group of series structures that are connected in parallel, as shown in Figure~\ref{fig:rbd_structures}(d). Its structure function can be expressed using OR of ANDs operators.

\section{Formalization of DRBDs in HOL}
\label{Formalization_in_hol}
%\vspace{-5pt}
In this section, we present our formalization for the proposed DRBD algebra including DRBD events, operators and constructs, simplification theorems and reliability expressions. First, we review some HOL probability theory preliminaries required for understanding the rest of the work.
%\vspace{-5pt}
\subsection{HOL Probability Theory}
%\vspace{-2pt}
\begin{table}[!b]
\centering
\setlength\tabcolsep{1pt}
%\vspace{-15pt}
\caption{HOL4 Probability Functions}
%%\vspace{-5pt}
\footnotesize
\label{table:assumption}
\begin{tabular}{|p{6.5cm}|p{9cm}|}
\hline
Function & Explanation \\ \hline \hline
{$\!\begin{aligned}[t]
	& {\texttt{rv\_gt0\_ninfinity L}}\end{aligned}$}                      
	
    & Random variables in list $L$ are greater than $0$ and not equal to $+\infty$   \\ \hline 

{$\!\begin{aligned}[t]
	& {\texttt{indep\_var p lborel}}\\[-2\jot]&{\texttt{~~(real o X) lborel (real o Y)}}\end{aligned}$}                           

& 
Independence of random variables defined from the probability space \texttt{p} to the Lebesgue Borel measure (\texttt{lborel})
             \\ \hline
{$\!\begin{aligned}[t]
	& {\texttt{distributed p lborel (real o X) f\textsubscript{x}}}\end{aligned}$}                           
& 
Defines a density function \texttt{f\textsubscript{x}} for the real version of random variable $X$ defined from the probability space $p$ to the Lebesgue-Borel measure\\ \hline

{$\!\begin{aligned}[t]
	&{\texttt{measurable\_CDF p (real o Y) }}\end{aligned}$}                           

& 
Ensures that CDF (F\textsubscript{Y}) is measurable \\ \hline

{$\!\begin{aligned}[t]
	& {\texttt{cont\_CDF p (real o Y) }}\end{aligned}$}                           

& 
Ensures that CDF (F\textsubscript{Y}) is continuous \\ \hline

{$\!\begin{aligned}[t]
	& {\texttt{cond\_density lborel lborel p }}\\[-2\jot]&{\texttt{~~(real o X)(real o Y) y f\textsubscript{xy} f\textsubscript{y} f\textsubscript{X\textsubscript{a}|Y} }}\end{aligned}$}                           

& 
Defines a conditional density function f\textsubscript{X\textsubscript{a}$|$Y} using the joint density function f\textsubscript{xy} and the marginal density function f\textsubscript{y} \\ \hline

{$\!\begin{aligned}[t]
	& {\texttt{den\_gt0\_ninfinity f\textsubscript{X\textsubscript{a}Y} f\textsubscript{Y} f\textsubscript{X\textsubscript{a}|Y}}}\end{aligned}$}                           

& 
Ensures the proper values for the density functions; joint, marginal and conditional, respectively. 0 $\leq$ f\textsubscript{X\textsubscript{a}Y}, 0 $<$ f\textsubscript{Y} and 0 $\leq$ f\textsubscript{X\textsubscript{a}$|$Y} \\ \hline

{$\!\begin{aligned}[t]
	& {\texttt{indep\_sets p X s}}\end{aligned}$}                           

& 
Ensures that the group of sets X indexed by the numbers in set s are independent over the probability space p \\ \hline
\end{tabular}
\end{table}
\indent The probability space is defined in HOL as a measure space, where the measure (probability) of the entire space is 1. A probability space is a measure space and is defined as a triplet $(\Omega, \mathcal{A}, \mathcal{P}r)$, where $\Omega$ is the space, $\mathcal{A}$ are the probability events and $\mathcal{P}r$ is the probability \cite{Mhamdi-entropy}. Two functions are defined in HOL; \texttt{p\_space p} and \texttt{events p}, that return the space ($\Omega$) of the above triplet and the events ($\mathcal{A}$), respectively.  A random variable is a measurable function that maps the probability space $p$ to another space. It is defined in HOL as \cite{Mhamdi-entropy}:

\begin{definition}
\label{DEF:random_variable}
{\small\textup{\texttt{$\vdash$ $\forall$X p s. random\_variable X p s $\Leftrightarrow$}\\
       \mbox {\texttt{prob\_space p $\wedge$ X $\in$ measurable (p\_space p, events p) s }}}}
\end{definition}

\noindent where \texttt{X} is the random variable, \texttt{p} is the probability space and \texttt{s} is the space that the random variable maps to. In our work, we use the \texttt{borel} space, which is defined over the real line~\cite{Qasim-CICM}. 

For a random variable $X$, the probability distribution is defined as the probability that this random variable belongs to a certain set \cite{Tarek-thesis}:

\begin{definition}
\label{DEF:distribution}
\emph{}\\
{\small\textup{\texttt{\texttt{$\vdash$ $\forall$p X. distribution p X = ($\lambda$s. prob p (PREIMAGE X s $\cap$ p\_space p)) }}}}
\end{definition}

The cumulative distribution function (CDF) is defined as \cite{elderhalli2019probabilistic}:
\begin{definition}
\label{DEF:Cumulative_density_function}
{\small\textup{\texttt{\texttt{$\vdash$ $\forall$p X t. CDF p X t = distribution p X \{y | y $\leq$ (t:real)\}} }}}
\end{definition}
\noindent where \texttt{p} is a probability space, \texttt{X} is a real-valued random variable and \texttt{t} is a variable of type real and represents time.

Independence of random variables is an important property that ensures that the probability of the intersection of the events of these random variables equals the product of the individual events. This definition is ported from Isabelle/HOL \cite{Isabelle} to HOL4 as \cite{Qasim-CICM}: 

 \begin{definition}
			\label{indep_vars}
{\small
				\textup{\texttt{$\vdash$ indep\_vars p M X ii = }\\
\mbox{\texttt{($\forall$i. i $\in$ ii $\Rightarrow$}}\\
\mbox{\texttt{~random\_variable (X i) p (m\_space (M i), measurable\_sets (M i))) $\wedge$}}\\
\mbox{\texttt{indep\_sets p }}\\
\mbox{\texttt{~($\lambda$i. \{PREIMAGE f A $\cap$ p\_space p|(f=X i) $\wedge$  A $\in$ measurable\_sets (M i)\}) ii}}}}

		\end{definition}

This definition ensures that a group $X$ is composed of random variables indexed by the elements in set $ii$ and that the events represented by the preimage of these random variables are independent using \texttt{indep\_sets}. \texttt{indep\_var} is defined, based on Definition \ref{indep_vars}, to capture the behavior of independence for two random variables \cite{Qasim-CICM}.

Finally, the Lebesgue integral is defined in HOL4 based on positive simple functions and then extended for positive functions and functions with positive and negative values \cite{Mhamdi-entropy}. Throughout this work, we use the Lebesgue integral for positive functions, i.e., \texttt{pos\_fn\_integral}, since we are integrating cumulative distribution and probability density functions, which are always positive. The integration is over the real line and thus we use the Lebesgue-Borel measure (\texttt{lborel}) \cite{Qasim-CICM} for this purpose. The boundaries of this integral can be identified using an indicator function by specifying the set of elements used in the integration. For example, $\int_{A} f dx$ can be represented as \texttt{pos\_fn\_integral lborel ($\lambda$x. indicator\_fn A * f x)}. However, for the ease of understanding, we use the regular mathematical expressions, i.e.,  we use $\int f\ dx$ to express integrals instead. Table \ref{table:assumption} lists the probability theory functions used in the rest of the work~\cite{elderhalli2019probabilistic}.
%\vspace{-5pt}
\subsection{DRBD Event}
%\vspace{-3pt}
\indent In our formalization, we define the inputs, or the random variables representing the time to failure of system components, as lambda abstracted functions with a return datatype of extended-real, which represents real numbers besides $\pm\infty$.  \\
\indent We define the DRBD event of Equation (\ref{eq:event}) as:
\begin{definition}
\label{DEF:DRBD_event}
\emph{}\\
{\small\textup{\texttt{\texttt{$\vdash$ $\forall$p X t. DRBD\_event p X t = \{s | Normal t < X s\} $\cap$ p\_space p}}}}
\end{definition}

\noindent where \texttt{Normal} typecasts the real value of \texttt{t} from real to extended-real. 
This type conversion is required since we need real-valued random variables. However, we need to deal with the extended-real datatype to model the \texttt{NEVER} element. Therefore, we define the time-to-failure functions to return extended-real and typecast the values from extended-real to real using the function \texttt{real} and vice versa using \texttt{Normal}.\\
\indent We define the reliability as the probability of the DRBD event according to Equation~(\ref{eq:rel}):
\begin{definition}
\label{DEF:Rel}
{\small\textup{\texttt{\texttt{$\vdash$ $\forall$p X t. Rel p X t = prob p (DRBD\_event p X t)}}}}
\end{definition}
We verify the relationship between the reliability and the CDF of Equation (\ref{eq:rel}) as:

\begin{theorem}
\label{thm:Rel-CDF}
{\small\textup{\texttt{\texttt{$\vdash$ $\forall$p X t. rv\_gt0\_ninfinity [X] $\wedge$}}}}\\
{\mbox{\textup{\texttt{~~random\_variable (real o X) p borel $\Rightarrow$}}}}\\
{\mbox{\textup{\texttt{~~(Rel p X t = 1- CDF p (real o X) t)}}}}
\end{theorem}
\noindent where \texttt{real} typecasts the values of the random variable from extended-real to real as the CDF is defined for real-valued random variables, \texttt{random\_variable (real o X) p borel} ensures that \texttt{(real o X)} is a random variable over the real line represented by the \texttt{borel} space, and \texttt{rv\_gt0\_ninfinity} ensures that the random variable is greater than or equal to $0$ and not equal to $+\infty$, as described in Table \ref{table:assumption}, which means that the time of failure of any component cannot be negative or $+\infty$. Theorem \ref{thm:Rel-CDF} is verified based on the fact that the \texttt{DRBD\_event} and the set of the CDF are the complement of each other. Therefore, the probability of one of them equals one minus the other. For the rest of the work, we will denote \texttt{CDF p (real o X) t} by $F_{X}(t)$ to facilitate the understanding of the theorems. 
\subsection{Identity Elements, Operators and Simplification Theorems}
Our formalization of the identity elements and the DRBD operators is listed in Table~\ref{table:element-operator}, where \texttt{extreal} is the extended-real datatype in HOL4, \texttt{PosInf} represents $+\infty$, \texttt{min} and \texttt{max} are HOL functions that return the minimum and  maximum values of their arguments, respectively. This formalization follows the proposed definitions in Tables \ref{table:or_and_reliability} and \ref{table:temporal_operators}. However, we define the operators as lambda abstracted functions to be able to conduct the probabilistic analysis later. In addition, we verify several simplification theorems based on the properties of \texttt{extreal} numbers in HOL and the definitions of the DRBD operators.
For example, the following theorem represents the distributive property of the after operator over the AND:
\begin{theorem}
{\small\textup{\texttt{$\vdash \forall$X Y Z. X $\rhd$ (Y $\cdot$ Z) = (X $\rhd$ Y)$\cdot$(X $\rhd$ Z)}}} 
\end{theorem}

Table \ref{table:simplification_theorems} lists the simplification theorems that we developed and verified in the proposed algebra.

\begin{table}[!t]
%\vspace{-10pt}
\caption{Definitions of Identity Elements and DRBD Operators}
\footnotesize
\centering
\label{table:element-operator}
\begin{tabular}{|l|l|p{7.5cm}|}
\hline
Element/Operator & Mathematical Expression  &  Formalization \\ \hline  \hline
{\footnotesize{\texttt{Always element}}} & 
 $\!\begin{aligned}[b]
		{\displaystyle ALWAYS\ =\ 0}
	\end{aligned}$&$\!\begin{aligned}[c]
	& \footnotesize{\texttt{$\vdash$ R\_ALWAYS = ($\lambda$s. (0:extreal))  }}\end{aligned}$  \\ \hline
{\footnotesize{\texttt{Never element}}} &	
$\!\begin{aligned}[b]
	{\displaystyle NEVER\ =\ \texttt{+$\infty$}}
	\end{aligned}$& $\!\begin{aligned}[c]
	& \footnotesize{\texttt{$\vdash$ R\_NEVER = ($\lambda$s. PosInf)  }}\end{aligned}$  \\ \hline
{\footnotesize{\texttt{AND}}}&
$\!\begin{aligned}[b]
	{{\displaystyle X \cdot Y= min (X ,Y)}}	\end{aligned}$& $\!\begin{aligned}[c]
	& \footnotesize{\texttt{$\vdash$ $\forall$X Y. 
		R\_AND X Y =($\lambda$s. min (X s) (Y s))}}\end{aligned}$   
 \\ \hline
{\footnotesize{\texttt{OR}}}&
$\!\begin{aligned}[b]
	{{\displaystyle X + Y= max (X, Y)}
}
	\end{aligned}$& $\!\begin{aligned}[c]
	& \footnotesize{\texttt{$\vdash$ $\forall$X Y. 
		R\_OR X Y = ($\lambda$s. max (X s) (Y s))
}}\end{aligned}$   
 \\ \hline

{\footnotesize{\texttt{After}}}&
$\!\begin{aligned}[b]
	{{\displaystyle X \rhd Y= }{\scriptsize
	\begin{cases}  X, &X > Y\\ +\infty, &X\leq Y
\end{cases}} 
}
	\end{aligned}$& $\!\begin{aligned}[c]
	& \footnotesize{\texttt{$\vdash$ $\forall$X Y. 
		R\_AFTER X Y =}}\\[-1\jot]
		&\footnotesize{\texttt{($\lambda$s. if Y s  < X s then X s else PosInf)
}}\end{aligned}$   
 \\ \hline
{\footnotesize{\texttt{Simultaneous}}}& $\!\begin{aligned}[b]
	{{\displaystyle X \Delta Y= }{\scriptsize
	\begin{cases}  X, &X = Y\\ +\infty, &X\neq Y
\end{cases}} 
}
	\end{aligned}$ & $\!\begin{aligned}[c]
	& \footnotesize{\texttt{$\vdash$ $\forall$X Y. 
		R\_SIMULT X Y =}}\\[-1\jot]
		&\footnotesize{\texttt{($\lambda$s. if X s  = Y s then X s else PosInf)
}}\end{aligned}$    \\ \hline
{\footnotesize{\texttt{Inclusive After}}}& $\!\begin{aligned}[b]
	{{\displaystyle X  \unrhd Y=}{\scriptsize
	\begin{cases}  X, &X \geq Y\\ +\infty, &X < Y
\end{cases}} 
}
	\end{aligned}$ & $\!\begin{aligned}[c]
	& \footnotesize{\texttt{$\vdash$ $\forall$ X Y. 
		R\_INCLUSIVE\_AFTER X Y =}}\\ 		&\footnotesize{\texttt{($\lambda$s. if Y s  $\leq$ X s then X s else PosInf)
}}\end{aligned}$    \\ \hline
\end{tabular}

\end{table}

\begin{table}[!b]
\caption{Formally Verified Simplification Theorems}
\small
\centering
\label{table:simplification_theorems}
\begin{tabular}{|l|}
\hline
\multicolumn{1}{|c|}{Simplification Theorem} \\ \hline
$\!\begin{aligned}[c]
	& \small{\texttt{$\vdash$ $\forall$X. ($\forall$s. 0 $\scriptstyle\leq$ X s) $\Rightarrow$ (X $\cdot$  R\_ALWAYS = R\_ALWAYS)}}
	\end{aligned}$          \\ \hline

$\!\begin{aligned}[c]
	& \small{\texttt{$\vdash$ $\forall$X Y Z. (X $\cdot$ Y) $\cdot$ Z = X $\cdot$ (Y $\cdot$ Z)}}
	\end{aligned}$          \\ \hline

$\!\begin{aligned}[c]
	& \small{\texttt{$\vdash$ $\forall$X Y. X $\cdot$ Y = Y $\cdot$ X}}
	\end{aligned}$          \\ \hline

$\!\begin{aligned}[c]
	& \small{\texttt{$\vdash$ $\forall$X. X $\cdot$ X = X}}
	\end{aligned}$          \\ \hline

$\!\begin{aligned}[c]
	& \small{\texttt{$\vdash$ $\forall$X. X $\cdot$ R\_NEVER = X}}
	\end{aligned}$          \\ \hline

$\!\begin{aligned}[c]
	& \small{\texttt{$\vdash$ $\forall$X. ($\forall$s. 0 $\scriptstyle\leq$ X s) $\Rightarrow$ (X +  R\_ALWAYS = X)}}
	\end{aligned}$          \\ \hline

$\!\begin{aligned}[c]
	& \small{\texttt{$\vdash$ $\forall$X Y Z. (X + Y) + Z = X + (Y + Z)}}
	\end{aligned}$          \\ \hline

$\!\begin{aligned}[c]
	& \small{\texttt{$\vdash$ $\forall$X Y. X + Y = Y + X}}
	\end{aligned}$          \\ \hline

$\!\begin{aligned}[c]
	& \small{\texttt{$\vdash$ $\forall$X. X + X = X}}
	\end{aligned}$          \\ \hline

$\!\begin{aligned}[c]
	& \small{\texttt{$\vdash$ $\forall$X. X + R\_NEVER = R\_NEVER}}
	\end{aligned}$          \\ \hline

$\!\begin{aligned}[c]
	& \small{\texttt{$\vdash$ $\forall$X Y. X + (X $\cdot$ Y) =X}}
	\end{aligned}$          \\ \hline

$\!\begin{aligned}[c]
	& \small{\texttt{$\vdash$ $\forall$X Y Z. X $rhd$ (Y $\rhd$ Z) = ((X $\rhd$ Y) + (X $\rhd$ Z)) (Y $\rhd$ Z)}}
	\end{aligned}$          \\ \hline

$\!\begin{aligned}[c]
	& \small{\texttt{$\vdash$ $\forall$X Y. (X $\rhd$ Y) + (Y $\rhd$ X) = R\_NEVER}}
	\end{aligned}$          \\ \hline

$\!\begin{aligned}[c]
	& \small{\texttt{$\vdash$ $\forall$X Y Z. X $\rhd$ (Y $\cdot$ Z) = (X $\rhd$ Y) $\cdot$ (X $\rhd$ Z)}}
	\end{aligned}$          \\ \hline

$\!\begin{aligned}[c]
	& \small{\texttt{$\vdash$ $\forall$X Y Z. X $\cdot$ (Y + Z) = (X $\cdot$ Y) + (X $\cdot$ Z)}}
	\end{aligned}$          \\ \hline

$\!\begin{aligned}[c]
	& \small{\texttt{$\vdash$ $\forall$X Y Z. X + (Y $\cdot$ Z) = (X + Y) $\cdot$ (X + Z)}}
	\end{aligned}$          \\ \hline

$\!\begin{aligned}[c]
	& \small{\texttt{$\vdash$ $\forall$X Y. X $\unrhd$ Y = (X $\rhd$ Y) $\cdot$ (X $\Delta$ Y)}}
	\end{aligned}$          \\ \hline

$\!\begin{aligned}[c]
	& \small{\texttt{$\vdash$ $\forall$X Y Z. X $\rhd$ (Y + Z) = (X $\rhd$ Y) + (X $\rhd$ Z)}}
	\end{aligned}$          \\ \hline

$\!\begin{aligned}[c]
	& \small{\texttt{$\vdash$ $\forall$X Y. X $\Delta$ Y = Y $\Delta$ X}}
	\end{aligned}$          \\ \hline
\end{tabular}
\end{table}

In order to verify the reliability of the DRBD constructs, such as the spare, we need first to verify the reliability of the DRBD operators that are used to express the structure function of these constructs. For the AND and OR operators, we verify their reliability expressions as in Theorems~\ref{thm-rel-and} and \ref{thm-rel-or}, respectively. 

\begin{theorem}
\label{thm-rel-and}
{\small\textup{\texttt{\texttt{$\vdash$ $\forall$p X t. rv\_gt0\_ninfinity [X;Y] $\wedge$ }}}}\\
\mbox{\small{\textup{\texttt{~~~~indep\_var p lborel (real o X) lborel (real o Y) $\Rightarrow$}}}}\\
\mbox{\small{\textup{\texttt{~~~~(Rel p (X$\cdot$Y) t = Rel p X t * Rel p Y t)}}}}
\end{theorem}

\begin{theorem}
\label{thm-rel-or}
{\small\textup{\texttt{\texttt{$\vdash$ $\forall$p X t. rv\_gt0\_ninfinity [X;Y] $\wedge$ }}}}\\
\mbox{\small{\textup{\texttt{~~~~indep\_var p lborel (real o X) lborel (real o Y) $\Rightarrow$}}}}\\
\mbox{\small{\textup{\texttt{~~~~(Rel p (X + Y) t = 1 - (1 - Rel p X t) * (1 - Rel p Y t))}}}}
\end{theorem}

We verify Theorem \ref{thm-rel-and} by first rewriting using Definition \ref{DEF:Rel}. Then, we prove that \texttt{DRBD\_event} of the AND operator equals the intersection of the individual events, as in Equation (\ref{and-intersect}). Utilizing the independence of the real-valued random variables \texttt{real o X} and \texttt{real o Y}, the probability of intersection of their events equals the product of the probability of the individual events. Since \texttt{X} and \texttt{Y} are greater than $0$ and are not equal to $+\infty$, based on the function \texttt{rv\_gt0\_ninfinity}, the events in the probability space that correspond to $X$ and $Y$ are equal to the ones that correspond to \texttt{real o X} and \texttt{real o Y}. As a result, the \texttt{DRBD\_events} of \texttt{X} and \texttt{Y} are independent. Hence, the probability of their intersection equals the product of the probability of the individual events, i.e., their reliability. Theorem~\ref{thm-rel-or} is verified in a similar way. However, we prove that the \texttt{DRBD\_event} of the OR operator equals the union of the individual events, as in Equation~(\ref{or-union}). We verify that this union of events equals to the complement of the intersection of the complements of the individual events. Now, Theorem~\ref{thm-rel-or} can be proven using the independence of random  variables.

We extend the definition of the AND and OR operators to n-ary operators, \texttt{nR\_AND} and \texttt{nR\_OR}, that can be used to represent the relationship between an arbitrary number of elements. We formally define n-ary AND (\texttt{nR\_AND}) as:

\begin{definition}
\label{def:nR_AND}
\emph{}\\
{\small	\textup{\texttt{$\vdash$ $\forall$X s. nR\_AND X s = ITSET ($\lambda$e acc. R\_AND (X e) acc) s R\_NEVER}}} 
\end{definition}

\noindent where \texttt{ITSET} is the HOL function to iterate over sets. This definition applies the \texttt{R\_AND} over the elements of \texttt{X} indexed by the numbers in \texttt{s}. \texttt{R\_NEVER} is the identity element of the \texttt{R\_AND} operator.

Similarly, we formally define n-ary OR (\texttt{nR\_OR}) as:

\begin{definition}
\label{def:nR_OR}
\emph{}\\
{\small	\textup{\texttt{$\vdash$ $\forall$X s. nR\_OR X s = ITSET ($\lambda$e acc. R\_OR (X e) acc) s R\_ALWAYS}}} 
\end{definition}

\noindent where \texttt{R\_ALWAYS} is the identity element of the \texttt{R\_OR} operator. 
The reliability of these two operators would be similar to the reliability of the series and parallel structures, respectively, as will be described in the following section.

Finally, we verify the reliability expression of the after operator utilizing our formalization in \cite{elderhalli2019probabilistic}, where the description of the assumptions is listed in Table \ref{table:assumption}:

\begin{theorem}
{\small	\textup{\texttt{$\vdash$ $\forall$X Y p f\textsubscript{x} t. rv\_gt0\_ninfinity [X; Y] $\wedge$ 0 $\leq$ t $\wedge$}}\\
{\mbox{\textup{\texttt{~~~~indep\_var p lborel (real o X) lborel (real o Y) $\wedge$}}}}\\
{\mbox{\textup{\texttt{~~~~distributed p lborel (real o X) f\textsubscript{x} $\wedge$ ($\forall$x. 0 $\leq$ f\textsubscript{x} x) $\wedge$}}}}\\
{\mbox{\textup{\texttt{~~~~cont\_CDF p (real o Y) $\wedge$ measurable\_CDF p (real o Y) $\Rightarrow$}}}}\\
{\mbox{\textup{\texttt{~~~~(Rel p (X$\rhd$Y) t = 1- $\int_{0}^{t}$f\textsubscript{X}(x) $\times$ F\textsubscript{Y}(x) $d$x)}}}}}
		\end{theorem}

The proof of this theorem is based on $Pr(Y < X < t) = \int_{0}^{t}f_{X}(x) \times F_{Y}(x)\ dx$, which has been verified in \cite{elderhalli2019probabilistic} using the properties of the Lebesgue integral and independence of random variables. The DRBD \textit{after} operator represents a situation where the system continues to work until two components fail in sequence.  Thus, the above expressions allow us to verify the reliability expression of the \textit{after} operator, as the DRBD and DFT events complement one another.

\subsection{DRBD Constructs and their Reliability Expressions}
As mentioned previously, the spare construct can have three variants according to the type of the spare block. We formally define the generic case, i.e., the warm spare (WSP) as:

\begin{definition}
\label{DEF:WSP}
{\small\textup{\texttt{\texttt{$\vdash$ $\forall$Y X\textsubscript{a} X\textsubscript{d}. R\_WSP Y X\textsubscript{a} X\textsubscript{d} = (X\textsubscript{a} $\rhd$ Y) $\cdot$ (Y $\rhd$ X\textsubscript{d}) }}}}
\end{definition}
Since the DRBD and DFT events complement one another, we use our formalization of the probability of failure of the warm spare gate \cite{elderhalli2019probabilistic} to verify the reliability of the WSP construct:

\begin{theorem}

			\label{thm:WSP_prob}

{\small
				\textup{\texttt{$\vdash$ $\forall$p Y X\textsubscript{a} X\textsubscript{d} t f\textsubscript{Y} f\textsubscript{X\textsubscript{a}Y} f\textsubscript{X\textsubscript{a}|Y}. 0 $\leq$ t $\wedge$}\\
{\mbox{\texttt{~~($\forall$s. ALL\_DISTINCT [X\textsubscript{a} s; X\textsubscript{d} s; Y s]) $\wedge$ DISJOINT\_WSP Y X\textsubscript{a} X\textsubscript{d} t $\wedge$}}}\\
{\mbox{\texttt{~~rv\_gt0\_ninfinity [X\textsubscript{a}; X\textsubscript{d}; Y]  $\wedge$ den\_gt0\_ninfinity f\textsubscript{X\textsubscript{a}Y} f\textsubscript{Y} f\textsubscript{X\textsubscript{a}|Y} $\wedge$}}}\\
{\mbox{\texttt{~~($\forall$y. cond\_density lborel lborel p (real o X\textsubscript{a})(real o Y) y f\textsubscript{X\textsubscript{a}Y} f\textsubscript{Y} f\textsubscript{X\textsubscript{a}|Y}) $\wedge$}}}\\
{\mbox{\texttt{~~indep\_var p lborel (real o X\textsubscript{d}) lborel (real o Y) $\wedge$}}}\\
{\mbox{\texttt{~~cont\_CDF p (real o X\textsubscript{d}) $\wedge$ measurable\_CDF p (real o X\textsubscript{d}) $\Rightarrow$}}}\\
{\mbox{\texttt{~~$\big($Rel p (R\_WSP Y X\textsubscript{a} X\textsubscript{d}) t) =}}}\\
{\mbox{\texttt{~~~1 - $(\int_{0}^{t} f\textsubscript{Y}(y) * (\int_{y}^{t}$ f\textsubscript{(X\textsubscript{a}|Y=y)}(x) dx$)$ dy + $\int_{0}^{t}$ f\textsubscript{Y}(y)F\textsubscript{X\textsubscript{d}}(y)dy$)\big)$
}}}}}
  				
\end{theorem}
\noindent where \texttt{ALL\_DISTINCT} ensures that the main and spare parts cannot fail at the same time, \texttt{DISJOINT\_WSP Y X\textsubscript{a} X\textsubscript{d} t} ensures that until time \texttt{t}, the spare can only fail in one of its states and \texttt{den\_gt0\_ninfinity} 
ascertains the proper values of the density functions; joint ($f_{XY}$), marginal ($f_{Y}$) and conditional ($f_{X_{a}|Y}$). The description of these assumptions is listed in Table \ref{table:assumption} \cite{elderhalli2019probabilistic}. More details about the formal definitions of these functions can be found in \cite{elderhalli2019probabilistic}.
Theorem \ref{thm:WSP_prob} is verified by first defining a conditional density function \texttt{f\textsubscript{X\textsubscript{a}|Y}} for random variables (\texttt{real o X\textsubscript{a}}) and (\texttt{real o Y}). This is required as the failure of the spare part is affected by the time of failure of the main part. Therefore, it is required to define this conditional density function then prove the expression based on the probability of failure of the DFT spare gate, which is verified based on the properties of the Lebesgue integral. 
We formally define the cold spare construct (CSP), which is a special case of the WSP, as:

\begin{definition}
\label{DEF:CSP}
{\small\textup{\texttt{\texttt{$\vdash$ $\forall$Y X. R\_CSP Y X = ($\lambda$s. if Y s < X s then X s else PosInf)}}}}
\end{definition}
This definition means that the CSP construct will continue to work until the main part fails then the spare part is activated and fails in its active state. It is worth noting that since the spare part has only one state that affects the behavior of the CSP, which is the active state, therefore, we do not use any subscript with the active state, as the dormant state has no effect here in the behavior. We verify the reliability of the CSP construct based on the probability of failure of the CSP gate as \cite{elderhalli2019probabilistic}:

\begin{theorem}

			\label{thm:CSP_prob}

{\small
				\textup{\texttt{$\vdash$ $\forall$p X Y f\textsubscript{XY} f\textsubscript{Y} f\textsubscript{X|Y} t. 0 $\leq$ t $\wedge$}\\
{\mbox{\texttt{~~rv\_gt0\_ninfinity [X; Y]  $\wedge$ den\_gt0\_ninfinity f\textsubscript{XY} f\textsubscript{Y} f\textsubscript{X|Y} $\wedge$}}}\\
{\mbox{\texttt{~~($\forall$y. cond\_density lborel lborel p (real o X)(real o Y) y f\textsubscript{XY} f\textsubscript{Y} f\textsubscript{X|Y}) $\wedge$}}}\\
{\mbox{\texttt{~~$\big($Rel p (R\_CSP Y X) t) =~1 - $(\int_{0}^{t} f\textsubscript{Y}(y) * (\int_{y}^{t}$ f\textsubscript{(X|Y=y)}(x) dx$)$ dy $\big)$
}}}}}
  				
\end{theorem}

The conditions required for this theorem are similar to the ones of Theorem \ref{thm:WSP_prob}, as the WSP exhibits the behavior of the CSP if the main part fails before the spare. 

Finally, we define the hot spare construct (HSP) as:

\begin{definition}
\label{DEF:HSP}
{\small\textup{\texttt{\texttt{$\vdash$ $\forall$Y X. R\_HSP Y X = ($\lambda$s. max (Y s) (X s))}}}}
\end{definition}

This means that the HSP acts like the OR operator, where at least one of the main or the spare parts should continue to work for the HSP construct to maintain its successful behavior. Therefore, we can use Theorem \ref{thm-rel-and} to express the reliability of the HSP construct.  

We formally define the series structure as:

\begin{definition}
\label{def:DRBD_parallel}
\small{\textup{\texttt{$\vdash \forall$Y s. DRBD\_series Y s  = $\displaystyle\bigcap_{i\in s}$ (Y i)}}}
\end{definition}

We define the series structure as a function that accepts a group of sets, \texttt{Y}, that are indexed by the numbers in set \texttt{s} and returns the intersection of these sets. 

The parallel structure is defined in a similar way but it returns the union of the sets rather than the intersection. We formally define it as:

\begin{definition}
\label{def:DRBD_parallel}
\small{\textup{\texttt{$\vdash \forall$Y s. DRBD\_parallel Y s  = $\displaystyle\bigcup_{i\in s}$ (Y i)}}}
\end{definition}

The group of sets, \texttt{Y}, in both structures represents a family of events, i.e, \texttt{Y} will be instantiated later with DRBD events. The reliability expressions of the series and parallel structures are given in Table \ref{table_DRBD_structure}. We verify these expressions as:

\begin{theorem}
\label{thm:rel_series}
\emph{}\\
\small{\textup{\texttt{$\vdash \forall$p X t s. s $\neq$ \{\} $\wedge$ FINITE s $\wedge$}}}\\\mbox{\small{\textup{\texttt{~indep\_sets p ($\lambda$i. \{rv\_to\_event p X t i\}) s $\Rightarrow$}}}}\\
\mbox{\small{\textup{\texttt{~(prob p (DRBD\_series (rv\_to\_event p X t) s) =}}}}\\
\mbox{\small{\textup{\texttt{~~Normal ($\displaystyle\prod_{i\in s}$ (real (Rel p (X i) t))))}}}}
\end{theorem}

\begin{theorem}
\label{thm:rel_parallel}
\emph{}\\
\mbox{\small{\textup{\texttt{$\vdash \forall$p X t s. s $\neq$ \{\} $\wedge$ FINITE s $\wedge$}}}}\\
\mbox{\small{\textup{\texttt{~indep\_sets p ($\lambda$i. \{rv\_to\_event p X t i\}) s $\wedge$}}}}\\
\mbox{\small{\textup{\texttt{~($\forall$i. i $\in$ s $\Rightarrow$ rv\_to\_event p X t i $\in$ events p) $\Rightarrow$}}}}\\
\mbox{\small{\textup{\texttt{~(prob p (DRBD\_parallel (rv\_to\_event p X t) s) =}}}}\\                         
\mbox{\small{\textup{\texttt{~~1 - Normal ($\displaystyle\prod_{i\in s}$ (real (1 - Rel p (X i) t))))}}}}
\end{theorem}

\noindent where \texttt{s$\neq$\{\} $\wedge$ FINITE s} ensures that the set of indices,  \texttt{s}, is nonempty and finite.  The reliability of the series structure is verified based on the independence of the input events using \texttt{indep\_sets}, which ensures that for the probability space \texttt{p}, the given group of sets (\texttt{($\lambda$i. \{rv\_ti\_event p X t i\}}) indexed by the numbers in set \texttt{s} are independent, as described in Table \ref{table:assumption}. The family of sets (\texttt{($\lambda$i. \{rv\_ti\_event p X t i\}}) represents the DRBD events of the group of time-to-failure functions, \texttt{X}. This is defined as:

\begin{definition}
\small{\textup{\texttt{$\vdash\forall$p X t. rv\_to\_event p X t = ($\lambda$i. DRBD\_event p (X i) t)}}} 
\end{definition}

The function \texttt{rv\_to\_event} enables us to create  the group of \texttt{DRBD\_event} of time-to-failure functions of system blocks (\texttt{X}).
Based on the independence of these sets and the definition of the series structure (intersection of sets), we verify that the probability of the series structure is equal to the product of the reliability of the individual blocks (\texttt{Rel p (X i) t}), where \texttt{i$\in$s}. The product function ($\scriptstyle\prod$) in HOL4 returns a real value and the probability returns \texttt{extreal}, therefore, it is required to typecast the product function to \texttt{extreal} using \texttt{Normal}. Similarly, the product function finds the product of real-valued functions, therefore, it is required to typecast the reliability function(\texttt{Rel}) to real using the \texttt{real} function. The parallel structure is verified in a similar way. We replace the parallel structure (the union of events) with the complement of the intersection of the complements of the events. Then, we verify that the probability of this complement equals one minus the probability of the intersection of the complements. This requires the added condition that all DRBD events created using \texttt{rv\_to\_event} belong to the events of the probability space \texttt{p}.  

In order to express the series and parallel structures using DRBD operators, we verify that these structures are equal to the DRBD events of the \texttt{nR\_AND} and \texttt{nR\_OR}, respectively:

\begin{theorem}
\label{thm:nR_AND_series}
{\textup{\small{\texttt{$\vdash\forall$p X t s.~FINITE s $\wedge$ s $\neq$ \{\} $\Rightarrow$}}}}\\
\mbox{\textup{\small{\texttt{~~~(DRBD\_event p (nR\_AND X s) t = DRBD\_series (rv\_to\_event p X t) s)}}}}
\end{theorem}

\begin{theorem}
\label{thm:nR_OR_parallel}
{\textup{\small{\texttt{$\vdash\forall$p X t s.~FINITE s $\wedge$ 0 $\leq$ t $\Rightarrow$}}}}\\
\mbox{\textup{\small{\texttt{~~~(DRBD\_event p (nR\_OR X s) t = DRBD\_parallel (rv\_to\_event p X t) s)}}}}
\end{theorem}

\noindent We verify Theorems \ref{thm:nR_AND_series} and \ref{thm:nR_OR_parallel} by inducting on set \texttt{s} using \texttt{SET\_INDUCT\_TAC} that will create two subgoals to be solved; one for the empty set and another one for inserting an element to a finite set. Furthermore, we use the fact that the DRBD event of the AND and OR operators equal the intersection and the union of the individual events, respectively. For Theorem \ref{thm:nR_OR_parallel}, an additional condition is required, \texttt{0$\leq$t}, to be able to manipulate the sets and reach the final form of the theorem.

Interestingly, these structures can be easily extended to model and verify more complex structures, such as two-level structures, i.e., series-parallel and parallel series structures. We formally verify the reliability of the series-parallel structure as:

\begin{theorem}
\label{thm:rel_series_parallel}
\mbox{\textup{\small{\texttt{$\vdash \forall$p X t s J.}}}}\\
\mbox{\textup{\small{\texttt{~indep\_sets p}}}}\\
\mbox{\textup{\small{\texttt{~~($\lambda$i. \{rv\_to\_event p X t i\}) ($\displaystyle\bigcup_{j\in J}$ (s j)) $\wedge$}}}}\\
\mbox{\textup{\small{\texttt{~($\forall$i. i $\in$ J $\Rightarrow$ s i $\neq$ \{\} $\wedge$ FINITE (s i)) $\wedge$}}}}\\
\mbox{\textup{\small{\texttt{~~FINITE J $\wedge$ J $\neq$ \{\} $\wedge$ disjoint\_family\_on s J $\Rightarrow$}}}}\\
\mbox{\textup{\small{\texttt{~(prob p}}}}\\
\mbox{\textup{\small{\texttt{~~~(DRBD\_series}}}}\\
\mbox{\textup{\small{\texttt{~~~~($\lambda$j. DRBD\_parallel}}}}\\
\mbox{\textup{\small{\texttt{~~~~~(rv\_to\_event p X t) (s j)) J) =}}}}\\
\mbox{\textup{\small{\texttt{~~Normal}}}}\\
\mbox{\textup{\small{\texttt{~~~($\displaystyle\prod_{j\in J}$ (1 -  $\displaystyle\prod_{i\in (s\ j)}$ (real (1 - Rel p (X i) t)))))}}}}
\end{theorem}

We formally verify the reliability of the parallel-series structure as:

\vspace{50pt}
\begin{theorem}
\label{thm:rel_parallel_series}
\emph{}\\
\mbox{\textup{\small{\texttt{$\vdash \forall$p X t s J.}}}}\\
\mbox{\textup{\small{\texttt{~indep\_sets p}}}}\\
\mbox{\textup{\small{\texttt{~~($\lambda$i. \{rv\_to\_event p X t i\}) ($\displaystyle\bigcup_{j\in J}$ (s j)) $\wedge$}}}}\\
\mbox{\textup{\small{\texttt{~($\forall$i. i $\in$ $\displaystyle\bigcup_{j\in J}$ (s j) $\Rightarrow$ rv\_to\_event p X t i $\in$ events p) $\wedge$}}}}\\
\mbox{\textup{\small{\texttt{~($\forall$i. i $\in$ J $\Rightarrow$ s i $\neq$ \{\} $\wedge$ FINITE (s i)) $\wedge$}}}}\\
\mbox{\textup{\small{\texttt{~FINITE J $\wedge$ J $\neq$ \{\} $\wedge$ disjoint\_family\_on s J $\Rightarrow$}}}}\\
\mbox{\textup{\small{\texttt{~(prob p}}}}\\
\mbox{\textup{\small{\texttt{~~(DRBD\_parallel}}}}\\
\mbox{\textup{\small{\texttt{~~~($\lambda$j. DRBD\_series}}}}\\
\mbox{\textup{\small{\texttt{~~~~(rv\_to\_event p X t) (s j)) J) =}}}}\\
\mbox{\textup{\small{\texttt{~~1 - }}}}\\
\mbox{\textup{\small{\texttt{~~Normal}}}}\\
\mbox{\textup{\small{\texttt{~~~($\displaystyle\displaystyle\prod_{j\in J}$ (1 - $\displaystyle\prod_{i \in (s\ j)}$ (real (Rel p (X i) t)))))}}}}
\end{theorem}

The main idea in building these two-level structures is to partition the family of blocks into distinct groups, where we use a set, \texttt{J}, to index these partitions, i.e., it includes the number of groups in the first top level. Then, for each group in this top level, we have another set,  \texttt{\{s j| j $\in$ J\}}, that includes the indices of the blocks in the second level, i.e. the subgroups. For example, consider the parallel-series structure of Figure~\ref{fig:rbd_structures}(d), if $n=m=1$, then the outer parallel structure has two series structures, where each series structure has two blocks. Thus, \texttt{J = \{0;1\}}. For each \texttt{j$\in$J}, we have a certain set \texttt{s j} that has the indices of the blocks in the inner series structure. Thus, \texttt{s = ($\lambda$j. if j = 0 then \{0;1\} else \{2;3\}}). The same concept is applied to the series-parallel structure. Therefore, the structure of the DRBD can be determined based on the given sets of indices.

We verify Theorems~\ref{thm:rel_series_parallel} and \ref{thm:rel_parallel_series} by extending the proofs of the series and parallel structures. However, it is required to deal with the intersection of unions in case of the series-parallel structure and the union of intersections in case of parallel-series structure. Therefore, we need to extend the independence of sets properties to include the independence of union and intersection of partitions of the events. We verify these properties as:

\begin{theorem}
\label{thm:indep_sets_collect_sigma_BIGUNION}
\small{\textup{\texttt{$\vdash \forall$p s J Y. indep\_sets p ($\lambda$i. \{Y i\}) $\bigcup_{j \in J}$ (s j) $\wedge$ J $\neq$ \{\} $\wedge$}}}\\
\mbox{\small{\textup{\texttt{~~($\forall$i. i $\in$ J $\Rightarrow$ countable (s i)) $\wedge$ FINITE J $\wedge$
          disjoint\_family\_on s J $\Rightarrow$}}}}\\
\mbox{\small{\textup{\texttt{~~indep\_sets p ($\lambda$j. \{$\bigcup_{i \in s\  j}$ (Y i)\}) J}}}}
\end{theorem}

\begin{theorem}
\label{thm:indep_sets_collect_sigma_BIGINTER}
\small{\textup{\texttt{\hspace{-2pt}$\vdash\forall$p s J Y. indep\_sets p ($\lambda$i. \{Y i\}) $\bigcup_{j \in J}$ (s j) $\wedge$ J $\neq$ \{\} $\wedge$}}}\\
\mbox{\small{\textup{\texttt{~~($\forall$i. i $\in$ J $\Rightarrow$ countable (s i) $\wedge$ s i $\neq$ \{\}) $\wedge$ FINITE J $\wedge$}}}}\\
\mbox{\small{\textup{\texttt{~~disjoint\_family\_on s J $\wedge$ ($\forall$i. i $\in$ $\bigcup_{i \in J}$ (s j) $\Rightarrow$ Y i $\subset$ m\_space p) $\Rightarrow$}}}}\\
\mbox{\small{\textup{\texttt{~~indep\_sets p ($\lambda$j. \{$\bigcap_{i \in s\  j}$ (Y i)\}) J}}}}
\end{theorem}

\noindent where set \texttt{J} includes the indices of the partitions and  \texttt{s} has the indices of the individual blocks of each partition, \texttt{disjoint\_family\_on} ensures that the indices of the blocks in different partitions are disjoint and \texttt{indep\_sets p ($\lambda$i. \{Y i\}) $\bigcup_{j \in J}$ (s j)} ensures the independence of the family of blocks \texttt{\{Y i\}} where the indices of the individual blocks are given by the union of \texttt{s}. In order to verify Theorems \ref{thm:indep_sets_collect_sigma_BIGUNION} and \ref{thm:indep_sets_collect_sigma_BIGINTER}, we need the fact that the $\sigma$-algebras generated by \texttt{($\lambda$j. $\bigcup_{i\in  s\ j}$\{Y i\}}) with index set \texttt{J} are independent. Then we verify that 
\texttt{$\forall$j. j $\in$ J, set \{$\bigcup_{i\in s\ j}$ \{Y i\}\}} is a subset of the $\sigma$-algebra generated by \texttt{$\bigcup_{i\in  s\ j}$\{Y i\}} . Finally, based on these intermediate verified steps and the definition of \texttt{indep\_sets}, we are able to verify these theorems.

In order to verify the reliability of the series-parallel structure, we need to ensure the independence of the individual blocks. Therefore it is required to combine the indices of all blocks into a single set using \texttt{$\scriptstyle\bigcup_{j\in J}$ (s j)} to be used with \texttt{indep\_sets}. To be able to use the reliability of the series structure in this proof, we use Theorem \ref{thm:indep_sets_collect_sigma_BIGUNION} to verify the independence of the unions of partitions of events. This means verifying that the parallel structures are independent, i.e., the probability of intersection of these parallel structures equals the product of the reliability of the parallel structures. Finally, several assumptions related to sets \texttt{\{s i| i $\in$ J\}} and \texttt{J} are required, which include that these sets are finite and nonempty. Finally, it is required that every block has a unique index, which is ensured using \texttt{disjoint\_family\_on}. The reliability of the parallel-series structure is verified in a similar manner based on the reliability of the parallel structure. We verify the independence of the intersection of partitions of events rather than the union using Theorem \ref{thm:indep_sets_collect_sigma_BIGINTER}. In addition, it is required that all DRBD events belong to the events of the probability space.

We extend the reliability of the two-level series-parallel structure to verify the reliability of a more nested structure, i.e., series-parallel-series-parallel, as:

\begin{theorem}
\label{thm:nested-series-parallel}
\textup{\small{\texttt{$\vdash\forall$p X t s L A J.}}}\\
\mbox{\textup{\small{\texttt{~($\forall$i. i $\in$ nested\_BIGUNION s L A J $\Rightarrow$~rv\_to\_event p X t i $\in$ events p) $\wedge$}}}}\\
\mbox{\textup{\small{\texttt{~indep\_sets p ($\lambda$i. \{rv\_to\_event p X t i\}) (nested\_BIGUNION s L A J) $\wedge$}}}}\\
\mbox{\textup{\small{\texttt{~sets\_finite\_not\_empty s L A J $\Rightarrow$}}}}\\
\mbox{\textup{\small{\texttt{~(prob p}}}}\\
\mbox{\textup{\small{\texttt{~~(DRBD\_series ($\lambda$j.}}}}\\
\mbox{\textup{\small{\texttt{~~~~DRBD\_parallel ($\lambda$a.}}}}\\
\mbox{\textup{\small{\texttt{~~~~~DRBD\_series ($\lambda$l.}}}}\\
\mbox{\textup{\small{\texttt{~~~~~~DRBD\_parallel (rv\_to\_event p X t) (s l)) (L a)) (A j)) J) =}}}}\\
\mbox{\textup{\small{\texttt{~~Normal}}}}\\
\mbox{\textup{\small{\texttt{~~~($\prod_{j\in J}$}}}}\\
\mbox{\textup{\small{\texttt{~~~~(1 - $\prod_{a\in (A\ j)}$(1 - $\prod_{l\in (L\ a)}$ (1 - $\prod_{i\in (s\ l)}$(real (1 - Rel p (X i) t)))))))}}}}
\end{theorem}
For this four-level nested structure, we have four sets (indexed sets) that determine the structure of the DRBD, which are: \texttt{J}, \texttt{A}, \texttt{L} and \texttt{s}. This is similar to the two-level nested structure but with a deeper hierarchy. Therefore, in order to combine the indices of all the individual blocks in the DRBD in a single set, we define \texttt{nested\_BIGUNION s L A J} to union the elements of all \texttt{s i}, where \texttt{i$\in$ L a}, \texttt{a$\in$ A j} and \texttt{j$\in$J}. This is done in a hierarchical manner and can be extended easily to deeper levels. We use the previously mentioned function to ensure that all the individual events belong to the probability events and are independent as well. Moreover, it is required to ensure that the sets are finite, disjoint and nonempty, just like the series-parallel structure. We combine these set-related conditions using the function \texttt{sets\_finite\_not\_empty}. Finally, we verify Theorem \ref{thm:nested-series-parallel} within two main steps. The first step is to verify the reliability of the outer series-parallel, which requires verifying the independence of the intersection of union of partition of the DRBD blocks, i.e., the inner series-parallel structures are independent. The second step is to verify the reliability of the inner series-parallel structures, which can be done based on some set manipulation. This theorem can be used to verify even deeper structures, which would require verifying the independence of more nested structures. We use Theorem \ref{thm:nested-series-parallel} to verify the reliability of the series-parallel-series structure as it represents a special case of the series-parallel-series-parallel, where each of the innermost parallel structures has only one block.  Our formalization follows the natural definitions of parallel and series structures. Moreover, our verified lemmas of independence allow verifying deeper structures, which makes our formalization flexible and applicable to model the most complex systems. The proof script, which is available at \cite{Yassmeen-DRBDcode}, of our formalization required around 3200 lines. In the following section, we utilize our formalization in the verification of the reliability of two real-world systems. 

\section{Applications}
\label{case_study}
%%\vspace{-2pt}

%\vspace{-5pt}
\begin{figure}[!t]
\centering
%\vspace{-20pt}
\includegraphics[scale=0.8]{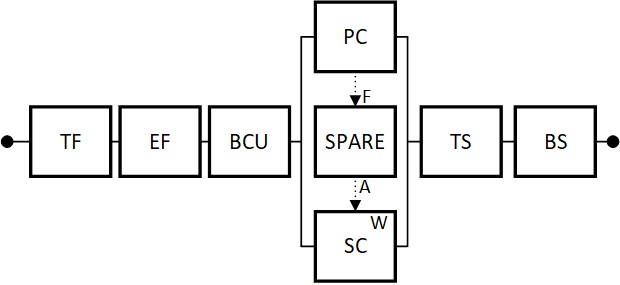}

\caption{DRBD of drive-by-wire system}
\label{fig:DBW}
%\vspace{-5pt}
\end{figure}
\begin{figure}[]
\centering
 \makebox[1\textwidth]{
\includegraphics[scale=0.7]{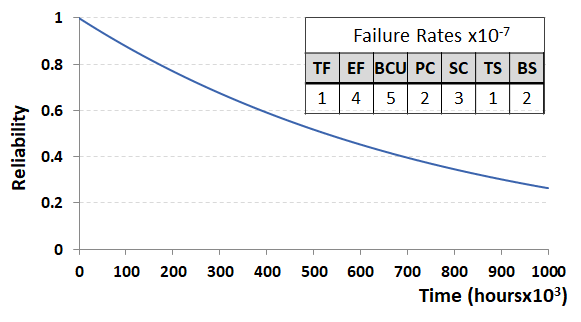}
}
\caption{Reliability of DBW system}
\label{plot_DBW}
\end{figure}

To demonstrate the applicability of our proposed DRBD algebra, we present the formal reliability analysis of a drive-by-wire system (DBW) \cite{altby2014design} and a shuffle-exchange network (SEN) \cite{bistouni2014analyzing} to verify generic expressions that are independent of the failure distribution of the system components, i.e., we can use different types of distributions to model the failure of system components as long as they satisfy the required conditions, such as the continuity.  \\
\indent The DRBD of the DBW system, shown in Figure~\ref{fig:DBW},
models the successful behavior of a key part of the modern automotive industry. This system controls the functionality of the
vehicle using a computerized controller. We provide the analysis of
the throttle and brake subsystems. The throttle subsystem continues to work as long as the throttle (TF) and the engine (EF) are working. In addition, the system successful operation requires the operation of the brake control unit
(BCU). The system includes a primary control
unit (PC) with a warm spare (SC) that replaces the main part after failure. Finally, the system needs the operation of the throttle sensor (TS) and the brake sensor (BS). The DRBD of this system is modeled as a series structure with a spare construct. We express the structure function of this DRBD using our operators:
\begin{equation*}
\small{\textup{\texttt{Q\textsubscript{DBW} = TF $\cdot$ EF $\cdot$ BCU $\cdot$ (R\_WSP PC SC\textsubscript{a} SC\textsubscript{d}) $\cdot$ TS $\cdot$ BS}}}
\end{equation*}
Then we verify the DBW reliability as:
\begin{thm}
\label{thm:Rel_DBW}
\textup{\small{\texttt{$\vdash\forall$p TF EF BCU PC SC\textsubscript{a} SC\textsubscript{d} TS BS t.}}}\\
\mbox{\textup{\small{\texttt{~DBW\_set\_req p TF EF BCU PC SC\textsubscript{a} SC\textsubscript{d} TS BS t $\Rightarrow$}}}}\\
\mbox{\textup{\small{\texttt{~(prob p (DRBD\_event p Q\textsubscript{DBW} t) =}}}}\\
\mbox{\textup{\small{\texttt{~~Rel p TF t * Rel p EF t * Rel p BCU t * Rel p (R\_WSP PC SC\textsubscript{a} SC\textsubscript{d}) t *}}}}\\ \mbox{\textup{\small{\texttt{~~Rel p TS t * Rel p BS t})}}}
\end{thm}
\noindent where \texttt{DBW\_set\_req} ensures the proper conditions for the independence of the blocks in the DBW system. In Figure~\ref{plot_DBW}, we evaluate, using MATLAB, the reliability of the DBW system assuming exponential distributions for the system components with failure rates as given in the figure and a dormancy factor of 0.5.

\begin{figure}[b]
%\vspace{-6pt}
\centering
{\includegraphics[scale=0.7]{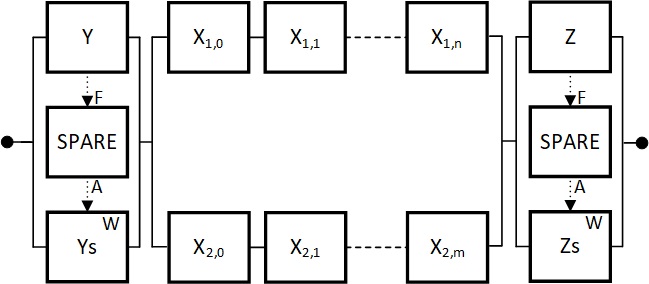}} 
%\vspace{-2pt}
\caption{DRBD of Shuffle-exchange Network with Spare Constructs}
\label{fig:SEN}
\end{figure}

In multi-processor systems, it is required to have an efficient communication method among system components, such as processors and memories. Multistage interconnection networks (MINs) can provide the necessary switching in multi-processor systems. A MIN consists of sources (inputs) and destinations (outputs) and is divided into a single-path MIN or a multiple-path MIN. In single-path MINs, there is only one possible path between each source and destination. Therefore, loosing any of the intermediate connections may lead to a failure. A SEN is an example of a single-path MIN. In order to increase the reliability of such network, additional switching elements are added to the network which provide additional paths between each source and destination. A SEN having two paths between each source and destination is usually called SEN+. The terminal reliability analysis, which is the reliability of the connection between a given source and destination, is usually conducted using traditional RBDs \cite{bistouni2014analyzing}. Although the reliability of the system is increased going from SEN to SEN+, each source is always connected to a single switch and the same thing applies to the destination. The failure of these single switches  would lead to the failure of the connection. Therefore, we propose to further enhance the reliability of this connection by using spare parts for these single switches so they can be replaced after failure. The DRBD of the modified SEN+ is shown in Figure \ref{fig:SEN}, where $Y$ and $Z$ are the main single switches that are connected to the source and destination with their spares $Ys$ and $Zs$, respectively. The parallel structure in the middle represents the reliability model of the two alternative paths between the source and the destination. Therefore, this DRBD consists of a series of two spare constructs and one parallel structure that consists of two series structures.

Using our DRBD operators, we formally express the structure function of this DRBD as:
%\vspace{-5pt}
\begin{equation}
\begin{split}
\small{\textup{\texttt{Q\textsubscript{SEN}}}} = &\small{\textup{\texttt{ nR\_AND ($\lambda$i. if i = 0 then R\_WSP Y Ys\textsubscript{a} Ys\textsubscript{d}}}}\\
&{\small{\textup{\texttt{~~~~~~~~~~~~~~else if i = 1 then \big((nR\_AND X L1) + (nR\_AND X L2)\big)}}}}\\
&{\small{\textup{\texttt{~~~~~~~~~~~~~~else R\_WSP Z Zs\textsubscript{a} Zs\textsubscript{d}) \{0; 1; 2\}}}}}
\end{split}
\end{equation}

Thus, the outer series structure is expressed using the \texttt{nR\_AND} operator over the set $\{0;1;2\}$ as this structure has three different structures; i.e., two spare constructs and one parallel structure. 
In order to re-utilize the verified expressions of reliability, it is required to express this DRBD using the series and parallel structures. Therefore, we verify that the DRBD event of the \texttt{Q\textsubscript{SEN}} is equal to a nested series-parallel-series structure as:

\begin{theorem}
\label{thm:SEN_nR_AND}
\textup{\small{\texttt{$\vdash\forall$p X Y Ys\textsubscript{a} Ys\textsubscript{d} Z Zs\textsubscript{a} Zs\textsubscript{d} t L1 L2.}}}\\
\mbox{\textup{\small{\texttt{~DISJOINT3 \{0; 3\} L1 L2 $\wedge$ FINITE L1 $\wedge$ FINITE L2 $\wedge$ L1 $\neq$ \{\} $\wedge$ L2 $\neq$ \{\} $\Rightarrow$}}}}\\
\mbox{\textup{\small{\texttt{~(DRBD\_event p Q\textsubscript{SEN} t =}}}}\\
\mbox{\textup{\small{\texttt{~~DRBD\_series ($\lambda$j.}}}}\\
\mbox{\textup{\small{\texttt{~~~~DRBD\_parallel ($\lambda$a.}}}}\\
\mbox{\textup{\small{\texttt{~~~~~~DRBD\_series ($\lambda$i.}}}}\\
\mbox{\textup{\small{\texttt{~~~~~~~~event\_set}}}}\\
\mbox{\textup{\small{\texttt{~~~~~~~~~[(DRBD\_event p (R\_WSP Y Ys\textsubscript{a} Ys\textsubscript{d}) t,0);}}}}\\
\mbox{\textup{\small{\texttt{~~~~~~~~~~(DRBD\_event p (R\_WSP Z Zs\textsubscript{a} Zs\textsubscript{d}) t,3)]}}}}\\
\mbox{\textup{\small{\texttt{~~~~~~~~~(rv\_to\_event p X t) i)}}}}\\
\mbox{\textup{\small{\texttt{~~~~~~~ind\_set [\{0\}; L1; L2; \{3\}] a))}}}}\\
\mbox{\textup{\small{\texttt{~~~~~(ind\_set [\{0\}; \{1; 2\}; \{3\}] j)) \{0; 1; 2\})}}}}
\end{theorem}

\noindent where \texttt{DISJOINT3} ensures that all sets are disjoint. Since \texttt{DRBD\_series} accepts a group of indexed sets, we define a function \texttt{event\_set} that accepts a list of pairs each of which is composed of a DRBD event with its index. This function also accepts the remaining blocks of the DRBD that have their indices embedded in a set (that can be generic of any size), such as the parallel structure of the SEN.  We also define \texttt{ind\_set} that accepts a list of sets and returns a group of indexed sets. Since we are dealing with a series-parallel-series structure, we need three sets to identify the hierarchy of this nested structure. Set $\{0;1;2\}$ in Theorem \ref{thm:SEN_nR_AND} indicates that the outer series structure has three elements, i.e., three parallel structures. \texttt{ind\_set [\{0\}; \{1;2\}; \{3\}]} indicates that the first parallel structure has only one series structure with index $0$, the second parallel structure has two series structures with indices $1$ and $2$, and the third parallel structure has only one series structure with index $3$. Finally, \texttt{ind\_set [\{0\}; L1; L2; \{3\}]} implies that the first series structure has only one element with index $0$, the second and third series structures have an arbitrary number of blocks indexed by $L1$ and $L2$. The last series structure has one element with index $3$. We verify Theorem \ref{thm:SEN_nR_AND} using Theorem \ref{thm:nR_AND_series} and the equivalence of the event of the OR with the union of events besides some set-related theorems. 

\begin{figure}[!t]
%\vspace{-20pt}
\centering
 \makebox[1\textwidth]{
{\includegraphics[scale=0.7]{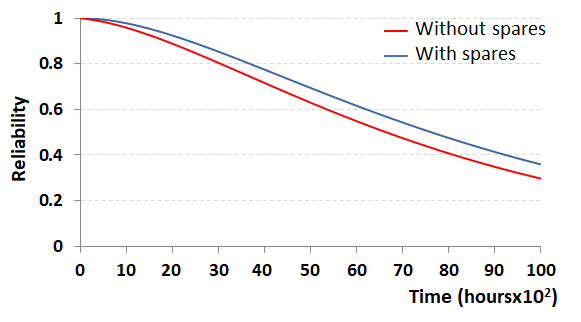}}} 
%\vspace{-2pt}
%\vspace{-7pt}
\caption{Reliability of SEN with/without spare constructs}
\label{plot}
%\vspace{-5pt}
\end{figure}

Based on Theorem \ref{thm:SEN_nR_AND}, we verify a generic expression for the reliability of the SEN system:
%%\vspace{-13pt}
\begin{theorem}
\label{thm:Rel_SEN}
\textup{\small{\texttt{$\vdash\forall$p X Y Ys\textsubscript{a} Ys\textsubscript{d} Z Zs\textsubscript{a} Zs\textsubscript{d} t L1 L2.}}}\\
\mbox{\textup{\small{\texttt{~SEN\_set\_req p L1 L2 (ind\_set [\{0\}; L1; L2; \{3\}])}}}}\\
\mbox{\textup{\small{\texttt{~~~(ind\_set [\{0\}; \{1; 2\}; \{3\}]) \{0; 1; 2\}}}}}\\
\mbox{\textup{\small{\texttt{~~~(event\_set [(DRBD\_event p (R\_WSP Y Ys\textsubscript{a} Ys\textsubscript{d}) t,0);}}}}\\
\mbox{\textup{\small{\texttt{~~~~~~~~~~~~~~~(DRBD\_event p (R\_WSP Z Zs\textsubscript{a} Zs\textsubscript{d}) t,3)] (rv\_to\_event p X t)) $\Rightarrow$}}}}\\
\mbox{\textup{\small{\texttt{~(prob p (DRBD\_event p Q\textsubscript{SEN} t) =}}}}\\
\mbox{\textup{\small{\texttt{~~Rel p (R\_WSP Y Ys\textsubscript{a} Ys\textsubscript{d}) t * Rel p (R\_WSP Z Zs\textsubscript{a} Zs\textsubscript{d}) t *}}}}\\
\mbox{\textup{\small{\texttt{~~(1 - (1 - Normal ($\prod_{l\in L1}$ (real (Rel p (X l) t)))) *}}}}\\
\mbox{\textup{\small{\texttt{~~~~~~~(1 - Normal ($\prod_{l\in L2}$ (real (Rel p (X l) t))))))}}}}
\end{theorem}

\noindent where \texttt{SEN\_set\_req} ensures the required conditions of the input sets including that the sets are finite and nonempty. It also ensures the independence of the input events over the probability space and that they belong to the probability events. We first rewrite the goal using Theorem \ref{thm:SEN_nR_AND}, then we use the reliability of the series-parallel-series to verify  the final expression. The reliability of the spare constructs can be further rewritten using Theorem \ref{thm:WSP_prob} given that the required conditions are ensured, such as the continuity of the CDFs. The final theorem with the expressions of the reliability of the spare constructs is available in \cite{Yassmeen-DRBDcode}.  Finally, we evaluate the reliability of the SEN system assuming the same failure rate of $1\times 10^{-5}$ for all switching elements. We also assume that each series structure has 16 switching elements. We evaluate the reliability for the SEN system without and with spare parts with a dormancy factor of 0.1, as shown in Figure~\ref{plot}. This result shows that considering the spares in the reliability analysis leads to having more reliable and realistic system than the traditional RBDs. 

To sum up, we are able to provide a generic expression of reliability of the SEN+ system that is verified in HOL theorem proving, which cannot be obtained using any other formal method. In addition, through the verified reliability expressions of the SEN+ and DBW systems, we demonstrated that our formalization is flexible and can be used to model more complex systems of an arbitrary number of blocks by implementing its hierarchy using sets that can be instantiated later to model a specific system structure, which is an added feature of our formalized algebra.
%\vspace{-20pt}
\section{Conclusion }
\label{Conclusion}
%\vspace{-4pt}
In this work, we proposed a new algebra to analyze dynamic reliability block diagrams (DRBDs). We developed the HOL formalization of this algebra in HOL4, which ensures its correctness and allows conducting the analysis within a theorem prover. Furthermore, this algebra provides formalized generic expressions of reliability that cannot be verified using other formal tools.  This HOL formalization is the first of its kind that takes into account the system dynamics by providing the HOL formal model of spare constructs and temporal operators. The proposed algebra is compatible with the reliability expressions of traditional RBDs as demonstrated by the reliability expressions of the series and parallel structures. It also facilitates extending the verified reliability expressions to model complex systems using nested structures.
Finally, we demonstrated the usefulness of this work by formally conducting the analysis of a drive-by-wire and a shuffle-exchange network systems to verify generic expressions of reliability, which are independent of the failure probability distribution of system components. We plan to extend this algebra to include other DRBD constructs, such as state dependencies, with their formalization in HOL, which would provide a more complete framework to algebraically analyze DRBDs within HOL theorem proving.

\bibliographystyle{unsrt}

\bibliography{mabiblio}
\end{document}